\documentclass[aps, prd, showpacs,amsmath,amssymb,superscriptaddress, nofootinbib, showkeys, notitlepage]{revtex4-2}

\usepackage[hyperindex]{hyperref}
\usepackage{amsfonts}
\usepackage{graphicx}
\usepackage{wasysym} 
\usepackage[makeroom]{cancel}
\usepackage{amsmath}
\usepackage{amssymb}
\usepackage{amsthm}
\usepackage{xcolor}
\usepackage{mathtools}
\usepackage{bbold}

\begin{document}
	
\title{Symplectic analysis for the Holst action with Dirac fields}

\author{Mariniel Galvão}
\email{mariniel.galvao@ufes.br}
\affiliation{Núcleo de Astrofísca e Cosmologia, PPGFis \& Departamento de Física,  Universidade Federal do Esp{\'i}rito Santo, Vit{\'o}ria, ES, Brazil}

\maketitle	

\section{Abstract}
	
In this work we review the canonical analysis of the Holst-Dirac action from the point of view of the Faddeev-Jackiw symplectic procedure using the Barcelo Neto-Wotzasek algorithm. We replicate the results found in the literature for the theory's constraints, recover some of the expressions for gauge symmetries of the main fields and perform the counting of the degrees of freedom. 
	
\section{Introduction}

The introduction of the tetrads formalism with Ashtekar variables has been fundamental for the research in quantum gravity theories since its inception\cite{Ashtekar:1986yd,Ashtekar:1987gu,Henneaux:1989hf}, as these variables make the Hamiltonian constraint polynomial in the new momenta and, consequently, easier to deal with. Besides, making use of the Ashtekar connection terms, the tetrad approach imbues the phase space in terms of a SU(2) gauge theory like Yang-Mills, useful in the development of a background independent formulation of General Relativity\cite{Ashtekar:1999dc}.

As already said, one of the merits of Ashtekar's formulation is to obtain a simple equation for the Hamiltonian constraint, in contrast to the complicated Wheeler-Dewit equation in the more traditional ADM formulation, although, as a result, complicated ``reality conditions'' appear and must be solved in order to obtain a real theory in terms of the Ashtekar variables with Lorentzian signatures\cite{Barbero:1994an,Barbero:1994ap}. This issue was tackled by Barbero through the introduction of self-dual $SO(4)$ connections leading to the $SO(3)$-ADM formalism, trivializing the reality condition equations, however this technique requires an appropriate choice for the Hamiltonian constraint, resulting in a more complicated expression. Both approaches were generalized and unified by the treatment conducted by Immirzi by the introduction of the $\beta$ parameter through the canonical transformation $\left( P^a_i, K^i_a \right) \rightarrow \left( P^a_i, A^i_a \eqqcolon \Gamma^i_a + \beta K^i_a \right) $ \cite{Immirzi:1996di,Immirzi:1996dr}. In his formulation, an imaginary $\beta$ is equivalent to Ashtekar's formulation, while $\beta \pm 1$ gives back Barbero's formulation.
	
Holst generalizes the previous results introducing the Ashtekar-Barbero-Immirzi Lagrangian, carried out from the Hilbert-Palatinin action, known as the Holst action\cite{Holst:1995pc}. Holst's modification is ``topological'' in vacuum, but not in the presence of fermions with minimal coupling\cite{Mercuri:2006um}. In this case there arises torsion terms in the II Cartan structure equation, as Bianchi's equation assumes the general form $R^a_b \wedge e^b = dT^a + \omega^a_b \wedge T^b$, which implies the presence of terms proportional to the Immirzi parameter in Einstein's field equations, so that General Relativity would no longer be obtained as the classical description. In fact, one hopes that it might be possible to find a Holst-Dirac type of action leading to an effective theory compatible with the Einstein-Cartan action when coupled to fermions. Motivated by this problem, Mercuri proposes a Holst action non-minimally coupled to fermionic fields in which the additional terms in relation to Einstein-Cartan's action result in a total divergence, equivalent to the topological invariant of Nieh-Yan\cite{Mercuri:2006um,Bojowald:2010qpa,Bojowald:2007nu}.

The symplectic analysis of gravitation theories with Immirzi parameters has been made in the context of Faddeev-Jackiw formalism, with the focus on obtaining the generalized brackets\cite{Escalante:2015wxi,Escalante:2016llg}, with results generally coinciding with those obtained in analysis via Dirac's formalism\cite{Sengupta:2009wg}, although some differences arise in regards to the set of constraints of the theory\cite{Escalante:2016llg}. There are also symplectic analysis of theories with half-integer spin variables not coupled to gravity\cite{Dengiz:2016xra}. But so far, to our knowledge, no systematic canonical analysis of gravitation theories coupled to fermions exist in the symplectic formalism. In this work such analysis is conducted through the Barcelos Neto-Wotzasek algorithm, replicating the results already found in the literature for the Holst-Dirac action, but in the light of symplectic geometrical framework. 
	
\section{Canonical Formulation of the Einstein-Cartan Action}\label{sc::TetradGravity}
Our entire analysis is a recreation of the work of Bojowald et al.\cite{Bojowald:2007nu} in the context of Faddeev-Jackiw's symplectic quantization. In this section and the next we recover some of the main results of the original paper while trying to highlight the main aspects of the theory from the point of view of symplectic geometrical analysis.
 
We start our analysis of Lorentzian gravity couple to Dirac fermions using the first order formalism for gravity, in this framework the field variables are the tetrad given by $e^I_\mu e^J_\nu \eta_{IJ} = g_{\mu \nu}$, a $so(3,1)$ connection 1-form, $\omega_{\mu \, J}^{\ I} = e^{\nu I}\nabla_\mu e_{\nu J}$, and the Dirac bi-spinor $\Psi = (\psi, \eta)^T$ and its complex conjugate $\bar{\Psi} = i \Psi^\dagger \gamma^0$, where $\gamma^\mu$ are the Dirac's gamma matrices. 

The Einstein-Cartan action minimally coupled to fermions is\cite{Mercuri:2006um,Bojowald:2007nu,Bojowald:2010qpa}
\begin{eqnarray}
	S[e,\omega,\Psi] = \frac{1}{2\kappa} \int_M d^4x \ |e| e^\mu_I e^\nu_J P^{IJ}_{\ \ \ KL} F^{KL}_{\mu \nu}(\omega) + \frac{1}{2} \int_M d^4x \ |e| \left[ \overline{\Psi} \gamma^I e_I^\mu \left( 1 - \frac{i}{\gamma}\gamma^5 \right) \nabla_\mu \Psi - \overline{\nabla_\mu \Psi} \left( 1 - \frac{i}{\gamma}\gamma^5 \right) \gamma^I e^\mu_I \Psi \right] \;, \nonumber \\
\end{eqnarray}
where $M : \mathcal{R} \times \Sigma_t$ represents the ADM-foliated space-time manifold, $F_{\mu \nu}^{IJ} = 2\partial_{[\mu} \omega_{\nu] \, J}^{\ I} + \omega_{[\mu}^{IK} \omega_{\nu]}^{LJ} \eta_{KL}$ and $\nabla_\mu \Psi = \partial_\mu \Psi + \frac{1}{4} \omega_\mu^{IJ} \gamma_{[I} \gamma_{J]} \Psi $ and $\gamma \in \mathbb{C}$ is the Immirzi parameter. Also, we have used 
\begin{eqnarray}
	P^{IJ}_{\ \ KL} = \delta^{[I}_K \delta^{J]}_L - \frac{1}{\gamma} \frac{\epsilon^{IJ}_{\ \ KL}}{2} \;,
\end{eqnarray}
where $\epsilon^{IJ}_{\ \ KL}$ is the totally anti-symmetric Levi-Civita tensor.

The variation of the action in respect to $\omega_\mu^{IJ}$ give equations which solutions yield the torsion contributions $C_{\mu I}^{\ J} v_J = \left( \nabla_\mu - \tilde{\nabla}_\mu \right) v_I$, where $\tilde{\nabla}_\mu$ is the covariant derivative compatible with the tetrad, which amounts to 
\begin{eqnarray} \label{eq::TorsionTerm}
	C_{aJK} = \frac{\kappa}{4} e^I_a \epsilon_{IJKL} J^L \;,
\end{eqnarray}
with $J^I = \overline{\Psi}\gamma^I\gamma_5\Psi$.

The spacetime foliation is done according to the prescription $t^\mu = Nn^\mu + N^\mu$, with $n_\mu N^\mu = 0$, where $N$ is called the lapse function, $N^\mu$ the shift vector and $n^\mu$ is the future pointing vector field normal to the surface $\Sigma_t$, i.e. $n^\mu n_\mu = -1$. The induced ``space'' metric induced $h_{\mu \nu}$ in the embedding $\Sigma$ is given by $g_{\mu \nu} = h_{\mu \nu} - n_\mu n_\nu$. By definition $h_{\mu \nu}n^\nu = 0$ and $N^\mu n_\mu = 0$, so we can use spatial indices $a,b,c,\ldots = 1,2,3$ for spatial tensor, e.g., $h_{ab}, N^a$ etc.

Additionally we use a partial gauge fixing convention on the internal indices, so we are able to split the tetrad into an internal time-like vector and a triad. Projecting the vierbein as 
$ h^\mu_\nu e^\nu_I \coloneqq \mathcal{E}^\mu_I = e^\mu_I + n^\mu n_I \;,$, where we used $n_I \eqqcolon e^\mu_I n_\mu$, satisfying $\mathcal{E}^\mu_I n_\mu = \mathcal{E}^\mu_I n^I = 0$. Choosing an internal time-like vector field $n^I = \delta_0^I$ we get the time-gauge description by requiring $e^\mu_0 = n^Ie^\mu_I = n^\mu$ is the unit normal to the foliation. Now, conveniently choosing the auxiliary variable $P^a_i \equiv \frac{\sqrt{h}}{\kappa\gamma} \mathcal{E}^a_i$ it is possible to find the canonical Holst-Dirac action in terms of canonical variables, 
\begin{eqnarray}\label{eq::AcaoCanonicaHolstDiracFinal}
	S_\text{H-D} &=& \int_{\mathcal{R}} dt \int_{\Sigma_t} d^3x \ \Bigg[ \dot{A}^i_a P^a_i + \dot{\psi} (i\sqrt{h}\theta_L\psi^\dagger) + \dot{\eta} (-i\sqrt{h}\theta_L\eta^\dagger) + \dot{\psi}^\dagger (-i\sqrt{h}\theta_R\psi) + \dot{\eta}^\dagger (i\sqrt{h}\theta_R\eta)  \nonumber \\
	&& + \Lambda^j \left( \mathcal{D}_b^{(A)}P^b_j - \frac{1}{2} J_i \right) + \omega_t^{0i}\left(  (1 + \gamma^2) \epsilon_{im}^{\ \ \ n} K_b^m P^b_n - \frac{\gamma^2 + 1}{2\gamma} J_i \right) \nonumber \\ 
	&& + N\left\{ \frac{\kappa}{2} \gamma^2 \frac{P^a_iP^b_j}{\sqrt{h}} \epsilon^{ij}_{\ \ k} \left[ \mathcal{F}^k_{ab} - 2(1 + \gamma^2) \epsilon^j_{\ qp} K_{a}^q K_{b}^p \right] + (1+\gamma^2)\kappa \tilde{D}_a \left(\frac{P^a_jG^j}{\sqrt{h}}\right) + \kappa (1 + \gamma^2) \frac{P^a_j}{2\sqrt{h}}\mathcal{D}^{(A)}_a(\sqrt{h}J^j) \right. \nonumber \\
	&& \left. + \gamma \kappa \frac{1 + \gamma^2}{2}\epsilon_{k \ m}^{\ l}K_a^kP_l^aJ^m + \gamma\kappa P^a_i \left[ i \theta_L ( \psi^\dagger \sigma^i \mathcal{D}^{(A)}_a \psi + \overline{\mathcal{D}^{(A)}_a \eta} \sigma^i \eta ) - i \theta_R ( \eta^\dagger \sigma^i \mathcal{D}^{(A)}_a \eta + \overline{\mathcal{D}^{(A)}_a \psi} \sigma^i \psi )  \right]  \right\} \nonumber \\
	&& - N^a \left\{ P^b_j \mathcal{F}_{ab}^j + \Big[ \theta_L (i \psi^\dagger \mathcal{D}_a^{(A)} \psi - i\overline{\mathcal{D}_a^{(A)} \eta} \eta) - \theta_R (i \overline{\mathcal{D}_a^{(A)}\psi} \psi - i \eta^\dagger \mathcal{D}_a^{(A)} \eta) \Big] - \frac{\gamma^2 + 1}{\gamma} K_a^j G_j  \right\} \Bigg] \;,
\end{eqnarray}
where $\theta_{L/R} := \frac{1}{2}( 1 \pm i/\gamma)$,
\begin{eqnarray}
	\mathcal{D}_a^{(A)} v_i &=& \nabla_a v^i + h_a^b \omega_{b \ j}^{\ i} v^i + 2\gamma h_a^b \omega_b^{\ 0k} v_k = \nabla_a v_i + \epsilon^i_{kl} \omega_a^{kl} + \gamma \epsilon_{ij}^{\ \ k} \omega_a^{\ 0j} v_k  \nonumber \\
	&=& \nabla_a v_i + \epsilon_{ij}^{\ \ k} \Gamma^j_a v_k + \gamma \epsilon_{ij}^{\ \ k} K_a^j v_k = \nabla_a v_i + \epsilon_{ij}^{\ \ k} A^j_a v_k  \\
	&=& \mathcal{D}_a v^i + \gamma \epsilon_{ij}^{\ \ k} K_a^j v_k   \;, \nonumber \\
	\mathcal{D}_a^{(A)} \Psi &=& \partial_a \Psi + \frac{1}{4} \omega_a^{ij} [\gamma_i, \gamma_ j] \Psi + i\frac{1}{4} \gamma K_a^i [\gamma_0, \gamma_i] = \mathcal{D}_a \Psi + i\frac{1}{4} \gamma K_a^i [\gamma_0, \gamma_i] \Psi \;, \\
	\Lambda^i &=& \frac{1}{2} \epsilon^j_{kl} \omega_t^{kl} + \gamma \omega_t^{0j}  \,.
\end{eqnarray}
And 
\begin{eqnarray}
	\mathcal{F}^l_{ab} = 2 \partial_{[a} \left( \Gamma_{b]}^l + \gamma K^l_{b]} \right) - \epsilon^l_{\ jk} \left( \Gamma_{[a}^j + \gamma K^j_{[a} \right) \left( \Gamma_{b]}^k + \gamma K^k_{b]} \right) = F^l_{ab} + 2\gamma \mathcal{D}_{[a} K^l_{b]} - \gamma^2 \epsilon^l_{\ jk} K^j_{[a} K^k_{b]} \;,
\end{eqnarray}
with $F_{ab}^l \equiv \frac{1}{2} \epsilon^l_{\ ij} F_{ab}^{ij} = 2\partial_{[a} \Gamma_{b]}^l - \epsilon^l_{\ jk} \Gamma_{a}^j \Gamma_{b}^k$ .
	
\section{Symplectic Analysis} \label{sc::EinsteiCartanAction}

Now the Barcelos Neto-Wotzasek algorithm for Grassmann variables (Appendix \ref{sec::MetSimpGrass}) is applied in order to conduct the symplectic analysis of the Holst-Dirac action. In the following analysis the constraints of the theory are obtained, spurious degrees of freedom are eliminated as some constraints are solved, and the remaining constraints are shown to generate gauge transformations, some of which are explicitly calculated, additionally the counting of the degrees of freedom is performed.
	
As the first step, we define the zeroth-order Lagrangian by
\begin{eqnarray} \label{eq::zeroOrderLagrangian}
	S_\text{H-D} &=& \int dt \ d^3x \ \mathcal{L}^{(0)} \;,
\end{eqnarray}
or
\begin{eqnarray}\label{eq::ZeroLagrangian}
	\mathcal{L}^{(0)} = \dot{A}^i_a P^a_i + \dot{\psi} (i\sqrt{h}\theta_L\psi^\dagger) + \dot{\eta} (-i\sqrt{h}\theta_L\eta^\dagger) + \dot{\psi}^\dagger (-i\sqrt{h}\theta_R\psi) + \dot{\eta}^\dagger (i\sqrt{h}\theta_R\eta)  - \mathcal{V}^{(0)} \;,
\end{eqnarray}
with the zeroth-order symplectic potential defined as
\begin{eqnarray}
	\mathcal{V}^{(0)} \equiv  \Lambda^i G_i + \omega_t^{\ 0i}S_i + NC + N^aC_a \;.
\end{eqnarray}

Following the BW algorithm, from (\ref{eq::zeroOrderLagrangian}), the zeroth-order symplectic vector and the zeroth-order symplectic one-form are written as
\begin{eqnarray}\label{eq::zerothOrderStuff}
	\begin{matrix}
	( \xi^{(0)} \,^{ \alpha } ) &=& 
	( A^i_a & P^a_i & \psi & \psi^\dagger & \eta & \eta^\dagger & \Lambda^i & K^i_a & \omega_t^{\ 0i}  & N & N^a )	 \\
	( a^{(0)} \,_{ \beta } ) &=& 
	( P^b_j & 0 & i\sqrt{h}\theta_L\psi^\dagger & -i\sqrt{h}\theta_R\psi & -i\sqrt{h}\theta_L\eta^\dagger & i\sqrt{h}\theta_R\eta & 0_j & 0_j^b & 0_j & 0 & \ 0_b \ \ )
	\end{matrix}
\end{eqnarray}

To write the zeroth-order symplectic vector the fields $\psi$, $\psi^\dagger$, $\eta$ e $\eta^\dagger$ were chosen as independent variables of the theory, likewise the corresponding one-forms for each of the fields were chosen as presented in the zeroth-order symplectic one-form. Such a choice is not free of problems, which will be discussed bellow.

From the zeroth-order symplectic objects (\ref{eq::zerothOrderStuff}), using Eq. (\ref{eq::sympMatrix}), we obtain the zeroth-order symplectic structure
\begin{eqnarray}
	f^{(0)}_{\alpha \beta}	&=& \frac{\delta a_{\beta}}{\delta \xi^{\alpha}} - \left( -1 \right)^{\varepsilon_\alpha \varepsilon_\beta} \frac{\delta a_{\alpha}}{\delta \xi^{\beta}} \nonumber \\
	&=& \delta_{\alpha}^2\delta_{\beta}^1 (\delta_j^i \delta^b_a) - \delta_{\alpha}^1\delta_{\beta}^2 (\delta_i^j \delta^a_b) + \delta_{\alpha}^4\delta_{\beta}^3 \left( \sqrt{h} \right) + \delta_{\alpha}^3\delta_{\beta}^4 \left( \sqrt{h} \right) + \delta_{\alpha}^6\delta_{\beta}^5 \left( \sqrt{h} \right) + \delta_{\alpha}^5\delta_{\beta}^6 \left( \sqrt{h} \right) \nonumber \\
	&& + \left( \delta_\alpha^2 \left[ \delta_\beta^3 \left( i\theta_L \psi^\dagger \right) + \delta_\beta^4 \left( -i\theta_R \psi \right) + \delta_\beta^5 \left( -i\theta_L \eta^\dagger \right) + \delta_\beta^6 \left( i\theta_R \eta \right)  \right] \frac{\delta \sqrt{h}}{\delta P^a_i} - \alpha \leftrightarrow \beta \right) \;. \label{eq::MatrizPS01}
\end{eqnarray}
Which the associated pre-symplectic matrix contains $5$ zero-modes on the symplectic space, namely,	
\begin{eqnarray}
	(\nu_1)^i &=& 
	\begin{pmatrix}
		0^i_a & 0^a_i & 0 & 0 & 1^i & 0^i_a & 0^i & 0 & 0^a
	\end{pmatrix}\;, \\
	(\nu_2)^i_a &=& 
	\begin{pmatrix}
		0^i_a & 0^a_i & 0 & 0 & 0^i & 1^i_a & 0^i & 0 & 0^a
	\end{pmatrix}\;, \\
	(\nu_3)^i &=& 
	\begin{pmatrix}
		0^i_a & 0^a_i & 0 & 0 & 0^i & 0^i_a & 1^i & 0 & 0^a
	\end{pmatrix}\;, \\
	(\nu_4) &=& 
	\begin{pmatrix}
		0^i_a & 0^a_i & 0 & 0 & 0^i & 0^i_a & 0^i & 1 & 0^a
	\end{pmatrix}\;, \\
	(\nu_5)^a &=& 
	\begin{pmatrix}
		0^i_a & 0^a_i & 0 & 0 & 0^i & 0^i_a & 0^i & 0 & 1^a
	\end{pmatrix}\;.
\end{eqnarray}

Applying the consistency conditions (\ref{eq::consistCond}) for each zero-mode, the following constraints are trivially obtained,
\begin{eqnarray}
	G^i &=& \mathcal{D}_b^{(A)}P^b_j - \frac{1}{2} J_i = 0 \;, \label{eq::GaussConstraint} \\
	\Omega^c_{l} &=& - \frac{(1+ \gamma^2)}{\gamma} \omega_t^{\ 0i} \epsilon_{il}^{\ \ m}P_m^c -2 \frac{(\gamma^2+1)}{\gamma^2} \epsilon_{lk}^{\ \ m} P^{[c}_m N^{a]} (A^k_a - \Gamma^k_a) - \frac{(\gamma^2+1)}{2 \gamma} \sqrt{h} J_l N^c \nonumber \\
	&& + \kappa N \left( \frac{1+\gamma^2}{\gamma} \right) P^c_l J_0 + N \mbox{sgn det}(e^i_a) \frac{(\gamma^2+1)}{\kappa \gamma} \sqrt{h} \Gamma^i_a e^a_{[i} e^c_{l]} + \frac{(\gamma^2+1)}{4\gamma \kappa} \varepsilon^{ac}_{\ \ d} \partial_{a} \left( Ne^d_l \right) \;, \label{eq::constraintwia} \\
	S_i &=& \mathcal{D}_b^{(A)}P^b_j - \frac{1}{2} J_i - (1 + \gamma^2) \epsilon_{im}^{\ \ \ n} K_b^m P^b_n + \frac{\beta}{2} J_i = 0 \;, \label{eq::SConstraint} \\
	C &=&  \frac{\kappa}{2} \gamma^2 \frac{P^a_iP^b_j}{\sqrt{h}} \epsilon^{ij}_{\ \ k} \left[ \mathcal{F}^k_{ab} - 2(1 + \gamma^2) \epsilon^j_{\ qp} K_{a}^q K_{b}^p \right] + (1+\gamma^2)\kappa \tilde{D}_a \left(\frac{P^a_jG^j}{\sqrt{h}}\right) + \kappa (1 + \gamma^2) \frac{P^a_j}{2\sqrt{h}}\mathcal{D}^{(A)}_a(\sqrt{h}J^j) \nonumber \\
	&& + \gamma \kappa \frac{1 + \gamma^2}{2}\epsilon_{k \ m}^{\ l}K_a^kP_l^aJ^m + \gamma\kappa P^a_i \left[ i \theta_L ( \psi^\dagger \sigma^i \mathcal{D}^{(A)}_a \psi + \overline{\mathcal{D}^{(A)}_a \eta} \sigma^i \eta ) - i \theta_R ( \eta^\dagger \sigma^i \mathcal{D}^{(A)}_a \eta + \overline{\mathcal{D}^{(A)}_a \psi} \sigma^i \psi ) \right] \;, \label{eq::vincHamiltonian01} \\
	C^a &=& P^b_j \mathcal{F}_{ab}^j + \Big[ \theta_L (i \psi^\dagger \mathcal{D}_a^{(A)} \psi - i\overline{\mathcal{D}_a^{(A)} \eta} \eta) - \theta_R (i \overline{\mathcal{D}_a^{(A)}\psi} \psi - i \eta^\dagger \mathcal{D}_a^{(A)} \eta) \Big] - \frac{\gamma^2 + 1}{\gamma} K_a^j G_j \;. \label{eq::diffConstraint}
\end{eqnarray}
	
Solving the Eq. (\ref{eq::constraintwia}) for $\omega_t^{\ 0i}$, we find the following expression for the Lagrange multiplier as a function of the triads,
\begin{eqnarray}
	\omega_t^{\ 0i} &=&  \frac{2}{3}\gamma \delta^{[i}_k \delta^{j]}_m e^{b}_j e^m_c N^{c} K^k_b + \frac{2}{3} \gamma N^{b} K^i_b + \frac{\kappa \gamma}{12} \epsilon^{il}_{\ \ m} J_l N^c e^m_c + \frac{\gamma}{12} e^{d[i}e^{al]} \partial_a (Ne_{dl}) + N\gamma \mbox{sgn det} (e^i_a) \epsilon^{il}_{\ \ m} e^a_{l} \Gamma^m_a \;.
\end{eqnarray}

From Eq. (\ref{eq::GaussConstraint}), we have that the constraint (\ref{eq::SConstraint}) can be written as
\begin{eqnarray}
	S_i = (1 + \gamma^2) \epsilon_{im}^{\ \ \ n} K_b^m P^b_n - \frac{\gamma^2 + 1}{2\gamma} J_i = 0 \label{eq::SConstraintFinal} \;,
\end{eqnarray}
which is equivalent to Eq. (\ref{eq::GaussConstraint}).

Now we will review the problems that arise from the choice of the fermionic variables in the composition of the symplectic space, how these problems present themselves in the symplectic framework, and how they are dealt with in the literature.

We need now to make a remark about Poisson brackets in the context of the symplectic formalism. Following \cite{Montani:1998ip,Montani:1993hf}, we have that, writing a simplified zeroth-order symplectic vector with only the variables appearing in the kinetic sector of $\mathcal{L}^{(0)}$, it is possible to write a generalized Poisson bracket given by
\begin{eqnarray}
	\{ E(\xi^{(0)}), G(\xi^{(0)}) \}_{\bar{f}} = \frac{\partial E}{\partial \xi^{(0)\alpha}} \left( \bar{f}^{(0) \alpha \beta} \right)^{-1} \frac{\partial G}{\partial \xi^{(0) \beta}} \;,
\end{eqnarray}
with $\bar{f}$ identified with Eq. (\ref{eq::sympMatrix}), deduced from Eq. (\ref{eq::0OLagrangian}). Explicitly, the invertible sector of the pre-symplectic matrix (\ref{eq::MatrizPS01}) is
\begin{eqnarray}
	\left( \bar{f}^{(0)}_{\alpha \beta} \right)  = 
	\begin{pmatrix}
	0 & -\delta^j_i \delta_b^a & 0 & 0 & 0 & 0 \\
	\delta_j^i \delta^b_a & 0 & i \theta_L \psi^\dagger \frac{\delta \sqrt{h}}{\delta P^a_i} & -i \theta_R \psi \frac{\delta \sqrt{h}}{\delta P^a_i} & -i \theta_L \eta^\dagger \frac{\delta \sqrt{h}}{\delta P^a_i} & i \theta_R \eta \frac{\delta \sqrt{h}}{\delta P^a_i} \\
	0 & -i \theta_L \psi^\dagger \frac{\delta \sqrt{h}}{\delta P^b_j} & 0 & \sqrt{h} & 0 & 0 \\
	0 & i \theta_R \psi \frac{\delta \sqrt{h}}{\delta P^b_j} & \sqrt{h} & 0 & 0 & 0 \\
	0 & \theta_L \eta^\dagger \frac{\delta \sqrt{h}}{\delta P^b_j} & 0 & 0 & 0 & \sqrt{h} \\
	0 & - i \theta_R \eta \frac{\delta \sqrt{h}}{\delta P^b_j} & 0 & 0 & \sqrt{h} & 0
	\end{pmatrix} \;.
\end{eqnarray}
Inverting the matrix above, we have
\begin{eqnarray}\label{eq::GeneralizePoissondBrackets}
	\left( \bar{f}^{(0) \alpha \beta} \right)^{-1} = 
	\begin{pmatrix}
	0 & \delta_j^i \delta^b_a & \frac{i \theta_R}{\sqrt{h}} \psi \frac{\delta \sqrt{h}}{\delta P^b_j} & -\frac{i \theta_L}{\sqrt{h}} \psi^\dagger \frac{\delta \sqrt{h}}{\delta P^b_j} & -\frac{i \theta_R}{\sqrt{h}} \eta \frac{\delta \sqrt{h}}{\delta P^b_j} & \frac{i \theta_L}{\sqrt{h}} \eta^\dagger \frac{\delta \sqrt{h}}{\delta P^b_j} \\
	-\delta^j_i \delta_b^a & 0 & 0 & 0 & 0 & 0 \\
	- \frac{i \theta_R}{\sqrt{h}} \psi \frac{\delta \sqrt{h}}{\delta P^a_i} & 0 & 0 & \frac{1}{\sqrt{h}} & 0 & 0 \\
	\frac{i \theta_L}{\sqrt{h}} \psi^\dagger \frac{\delta \sqrt{h}}{\delta P^a_i} & 0 & \frac{1}{\sqrt{h}} & 0 & 0 & 0 \\
	\frac{i \theta_R}{\sqrt{h}} \eta \frac{\delta \sqrt{h}}{\delta P^a_i} & 0 & 0 & 0 & 0 & \frac{1}{\sqrt{h}} \\
	-\frac{i \theta_L}{\sqrt{h}} \eta^\dagger \frac{\delta \sqrt{h}}{\delta P^a_i} & 0 & 0 & 0 & \frac{1}{\sqrt{h}} & 0
	\end{pmatrix} \;,
\end{eqnarray}
where we committed the indices on the zeros for simplicity. 

The generalized Poisson brackets given by Eq. (\ref{eq::GeneralizePoissondBrackets}) are problematic from the point of view of quantum mechanics. Because the fields that make the phase-space must, further down the road, be promoted to operators in a Hilbert space, the generalized Poisson brackets between the fermionic fields and the Ashtekar-Barbero connections must be zero, but, e.g. $\{ \psi, A^a_i \}_{\bar{f}} = - \frac{i\theta_R}{\sqrt{h}} \frac{\delta \sqrt{h}}{\delta P^a_i}$. Hence, the choice of variables for the symplectic space is deemed inadequate for the purpose of quantizing the theory and must be rectified.

Alternatively to the linearized Lagrangian (\ref{eq::ZeroLagrangian}), we may set the fields $\psi$ and $\eta$ as independent variables and use their canonical momenta. Then
\begin{eqnarray}
	\pi_\psi &=& \frac{\delta \mathcal{L}}{\delta \dot{\psi}} = i\sqrt{h}\theta_L \psi^\dagger \;, \\
	\pi_\eta &=& \frac{\delta \mathcal{L}}{\delta \dot{\eta}} = -i \sqrt{h} \theta_L \eta^\dagger \;.
\end{eqnarray}
Notice $\pi_{\overline{\psi}} = \overline{\pi_\psi}$ and $\pi_{\overline{\eta}} = \overline{\pi_\eta}$. However, these canonical variables also cannot be promoted to operators in a Hilbert space because they are proportional to $\sqrt{h}$ \cite{Thiemann:1997rq}. Besides, the kinetic sector of the action (\ref{eq::AcaoCanonicaHolstDiracFinal}) may be written in this case as 
\begin{eqnarray}
	\Theta = i\int_{\Sigma_t} d^3x \ \sqrt{h} \left( \theta_L \psi^\dagger \dot{\psi} - \theta_R \dot{\psi}^\dagger \psi - \theta_L \eta^\dagger \dot{\eta} + \theta_R \dot{\eta}^\dagger \eta \right) = \int_{\Sigma_t} d^3x \ \sqrt{h} \left[ \pi_\psi \dot{\psi} + \pi_\eta \dot{\eta} - \frac{i}{2}\gamma\kappa \theta_R e^i_c J^0 \dot{P}^c_i \right] \;,
\end{eqnarray}
aside from surface terms. And the second term of the integral contributes an imaginary correction to the connection term $A^i_a$, which is also undesirable. 

To solve this problem and find a real Ashtekar-Barbero connection for an action with fermionic contributions, it suffices to redefine the fermion fields into Grassmann variables of density one-half. In other words, we define 
\begin{eqnarray}
	\xi &\coloneqq& \sqrt[4]{h}\psi \;, \label{eq::FermionXi} \\
	\chi &\coloneqq& \sqrt[4]{h}\eta \label{eq::FermionChi}
\end{eqnarray}
as the new canonical variables, and
\begin{eqnarray}
	\pi_\xi &=& -i\xi^\dagger \;, \label{eq::FermionMomentumXi} \\
	\pi_\chi &=& -i\chi^\dagger \label{eq::FermionMomentumChi}
\end{eqnarray}
as their canonical conjugate momenta. In this way the symplectic term in the action (\ref{eq::AcaoCanonicaHolstDiracFinal}), may be written as
\begin{eqnarray}
	\Theta_\text{Dirac} &=& -\frac{i}{2} \int d^3x \sqrt{h} \left[ \left( 1 + \frac{i}{\gamma} \right) \left( \psi^\dagger \dot{\psi} - \dot{\eta^\dagger} \eta \right) - \left( 1 - \frac{i}{\gamma} \right) \left( \dot{\psi^\dagger} \psi - \eta^\dagger \dot{\eta} \right) \right]  \nonumber \\
	&=&  \int d^3x  \left[ -i\left( \xi^\dagger \dot{\xi} + \chi^\dagger \dot{\chi}  \right) + \frac{\kappa}{4} P^a_i \mathcal{L}_t ( e^i_a J^0 ) \right] \nonumber \\
	&=&  \int d^3x  \left[  \pi_\xi \dot{\xi} + \pi_\chi \dot{\chi} + \frac{\kappa}{4} P^a_i \mathcal{L}_t ( e^i_a J^0 ) \right] \;. \label{eq::TermoSimpleticoPsiDagaPsi}
\end{eqnarray}
And the kinetic sector of the Holst-Dirac action may be rewritten as
\begin{eqnarray} \label{eq::AcaoHolstDiracCanonicaAlt}
	S_{\text{H-D}} &=& \int_\mathbb{R} dt \int_\Sigma d^3x \left\{  \pi_\xi \dot{\xi} + \pi_\chi \dot{\chi} +  P^a_i \mathcal{L}_t \left( \frac{\kappa}{4} e^i_a J^0 + A^i_a \right) \right\} \;.
\end{eqnarray}
Hence, given the redefinitions of the fermions into half-densities, the corrected Ashtekar-Barbero connection is
\begin{eqnarray}
	\mathcal{A}^i_a \coloneqq A^i_a + \frac{\kappa}{4} e^i_aJ^0 \;.
\end{eqnarray}
	
Additionally, taking the Gaussian constraint (\ref{eq::GaussConstraint}) plus the constraint (\ref{eq::SConstraintFinal}), we have
\begin{eqnarray} \label{eq::DerPZero}
	\mathcal{D}_a^{(A)} P^a_i - \frac{\sqrt{h}}{2}J_i = \mathcal{D}_a P^a_i + \gamma \epsilon_{ij}^{\ \ k} K_a^j P^a_k - \frac{\sqrt{h}}{2}J_i = \mathcal{D}_a P^a_i = 0 \;,
\end{eqnarray}
even in the presence of fermions, so
$
	\epsilon^{ij}_{\ \ k} \Gamma^k_b P^b_j = \epsilon^{ij}_{\ \ k} \tilde{\Gamma}^k_b P^b_j \;,
$
everywhere, what implies that
\begin{eqnarray} \label{eq::PbCbCContraction}
	\epsilon^{ij}_{\ \ k} C^k_b P^b_j = 0 
\end{eqnarray}
on the constraint surface given by $S^i = 0$. Which agrees with the description of the theory endowed with a Lorentz connection. Spatially projecting the space-temporal contribution of the torsion term $C_\mu^{IJ}$ given by Eq. (\ref{eq::TorsionTerm}), we have the contribution in terms of the triads as
\begin{equation}
	C^j_a = \frac{1}{2} h^b_a \epsilon^{IJ}_{\ \ KL} n_I C_b^{KL} = -\frac{\kappa}{4}e^j_aJ^0\;.
\end{equation}
Yielding the Levi-Civita connection
\begin{eqnarray}
	\Gamma^k_b = \tilde{\Gamma}^k_b - \frac{\kappa}{4} e^k_b J^0 \;,
\end{eqnarray}
as a sum of a torsionless term and a torsion contribution.
	
So the Ashtekar-Barbero may be written as
\begin{eqnarray}
	A^i_a &=& \tilde{\Gamma}^i_a + \gamma K^i_a - \frac{\kappa}{4} e^i_aJ^0 
\end{eqnarray}
and
\begin{eqnarray}
	\mathcal{D}_a^{(A)} P^a_i &=& \mathcal{D}_a^{(\mathcal{A})} P^a_i \coloneqq \mathcal{D} P^a_i + \epsilon_{ij}^{\ \ k} \mathcal{A}_a^j P^a_k \;.
\end{eqnarray}
From this result we define the corrected Ashtekar-Barbero connection as
\begin{eqnarray}
	\mathcal{A}^i_a = \tilde{\Gamma}^i_a + \gamma K^i_a \;.
\end{eqnarray}

Bellow we rewrite the constraints of the theory using the half-densitized fermionic variables and the corrected Ashtekar-Barbero connection. Starting by the Gaussian constraint, we have
\begin{eqnarray}
	\mathcal{D}_b^{(A)} P^b_i - \frac{\sqrt{h}}{2} J_i &=&  \mathcal{D}_b^{(\mathcal{A})} P^b_i - \frac{\sqrt{h}}{2} \left( \psi^\dagger \sigma_i \psi + \eta^\dagger \sigma_i \eta  \right) 
\end{eqnarray}
or
\begin{eqnarray}
	G_i &=&  \mathcal{D}_b^{(\mathcal{A})} P^b_i + \left( \pi_\xi \tau_i \xi + \chi \tau_i \chi  \right) \label{eq::GaussConstraintNew} \;,
\end{eqnarray}
with $\tau_i \coloneqq \frac{\sigma}{2i}$.

Next we have the Hamiltonian constraint in the new coordinates,
\begin{eqnarray}
	\mathcal{C} &=& \frac{\gamma^2 \kappa}{2\sqrt{h}}  P^a_i P^b_j \left( \epsilon^{ij}_{\ \ k} \mathcal{F}^{(\mathcal{A})k}_{\ \ ab} - 2(\gamma^2+1)K^i_{[a}K^j_{b]} \right) - \frac{\kappa (\gamma^2+1)}{\sqrt{h}}P^a_i \mathcal{D}_a^{(\mathcal{A})} ( \pi_\xi
	\tau_i \xi + \pi_\chi \tau_i \chi ) \nonumber \\
	&& -i\frac{2\gamma \kappa}{\sqrt{h}} P^a_i \left( \theta_L \pi_\xi \tau^i \mathcal{D}_a^{(\mathcal{A})} \xi - \theta_R \pi_\chi \tau^i \mathcal{D}_a^{(\mathcal{A})} \chi - c.c. \right) + i\frac{\gamma^2\kappa^2}{4\sqrt{h}} \epsilon^{ij}_{\ \ k} P^a_i e^k_b \left( \pi_\xi \xi - \pi_\chi \chi \right) \partial_{a} P^b_j \nonumber \\
	&& + \frac{\kappa}{4 \sqrt{h}}(\gamma^2+ 1)  \left( \pi_\xi \tau_j \xi + \pi_\chi \tau_j \chi \right) \left( \pi_\xi \tau^j \xi + \pi_\chi \tau^j \chi \right) \;. \label{eq::HamiltonianConstraintNew}
\end{eqnarray}
	
Finally, the expression for the spatial diffeomorphism constraint in the new coordinates is
\begin{eqnarray}
	\mathcal{C}_a &=& \left( 2P^b_k \partial_{[a} \mathcal{A}_{b]}^k - \mathcal{A}^k_a \partial_b P^b_k \right) + A^k_a \left( \mathcal{D}^{(A)}_b P^b_k - \frac{\sqrt{h}}{2} J_k  \right) \nonumber \\ 
	&& +  \frac{1}{2}  \left( \pi_\xi \partial_a \xi - (\partial_a \pi_\xi) \xi -  \pi_\chi \partial_a \chi  +  (\partial_a \pi_\chi) \chi \right) + P^b_k \partial_a C_b^k - \frac{\sqrt{h}\kappa}{4} P^b_j \partial_a C_b^j \nonumber \\
	&=&  \ \left( 2P^b_k \partial_{[a} \mathcal{A}_{b]}^k - \mathcal{A}^k_a \partial_b P^b_k \right) + \frac{1}{2}  \left( \pi_\xi \partial_a \xi - (\partial_a \pi_\xi) \xi -  \pi_\chi \partial_a \chi  +  (\partial_a \pi_\chi) \chi \right) \;. \label{eq::DiffConstraintNew}
\end{eqnarray}

Incorporating the constraints into the symplectic space of the next BW iteration and using the new variables we write the first-order symplectic Lagrangian as
\begin{eqnarray}
	\overset{(1)}{\mathcal{L}} = P^a_i \dot{\mathcal{A}}^i_a + \pi_\xi \dot{\xi} + \pi_\chi \dot{\chi} + \dot{\lambda}^i \mathcal{G}_i + \dot{\lambda} \mathcal{C} + \dot{\lambda}^a \mathcal{C}_a \;,
\end{eqnarray}
with first-order symplectic vector and 1-form given, respectively, by
\begin{eqnarray}
	( \overset{(1)}{\xi} \,^{ \alpha } ) &=& 
	\begin{pmatrix}
	\mathcal{A}^i_a & P^a_i & \xi & \pi_\xi & \chi & \pi_\chi & \lambda^i & \lambda & \lambda^a 
	\end{pmatrix}\;. \label{eq::sympVector01} \\
	( \overset{(1)}{a} \,_{ \beta } ) &=& 
	\begin{pmatrix}
	P^b_j & 0_b^j & \pi_\xi & 0 & \pi_\chi & 0 & \mathcal{G}_j & \mathcal{C} & \mathcal{C}_b
	\end{pmatrix}\;.
\end{eqnarray}
And the first-order pre-symplectic structure is, thus,
\begin{eqnarray}
	\overset{(1)}{f}_{\alpha \beta}	&=& \frac{\delta a_{\beta}}{\delta \xi^{\alpha}} - \left( -1 \right)^{\varepsilon_R\varepsilon_S} \frac{\delta a_{\alpha}}{\delta \xi^{\beta}} \nonumber \\
	&=& \delta_\alpha^2\delta_\beta^1 \delta_b^a \delta^j_i - \delta_\alpha^1\delta_\beta^2 \delta_a^b \delta^i_j + \delta_\alpha^4\delta_\beta^3 + + \delta_\alpha^3\delta_\beta^4 + \delta_\alpha^6\delta_\beta^5 + \delta_\alpha^5\delta_\beta^6 \nonumber \\
	&& +\left[ \left( \delta_\alpha^1 \frac{\delta}{\delta \mathcal{A}^i_a} + \delta_\alpha^2 \frac{\delta}{\delta P^a_i} + \delta_\alpha^3 \frac{\delta}{\delta \xi} + \delta_\alpha^4 \frac{\delta}{\delta \pi_\xi} + \delta_\alpha^5 \frac{\delta}{\delta \chi} + \delta_\alpha^6 \frac{\delta}{\delta \pi_\chi} \right) \mathcal{G}_j \delta_\beta^7 -\left( -1 \right)^{\varepsilon_R\varepsilon_S}  \alpha \leftrightarrow \beta \right] \nonumber \\
	&& + \left[ \left( \delta_\alpha^1 \frac{\delta}{\delta \mathcal{A}^i_a} + \delta_\alpha^2 \frac{\delta}{\delta P^a_i} + \delta_\alpha^3 \frac{\delta}{\delta \xi} + \delta_\alpha^4 \frac{\delta}{\delta \pi_\xi} + \delta_\alpha^5 \frac{\delta}{\delta \chi} + \delta_\alpha^6 \frac{\delta}{\delta \pi_\chi} \right) \mathcal{C} \delta_\beta^8 -\left( -1 \right)^{\varepsilon_R\varepsilon_S}  \alpha \leftrightarrow \beta \right] \nonumber \\
	&& +\left[ \left( \delta_\alpha^1 \frac{\delta}{\delta \mathcal{A}^i_a} + \delta_\alpha^2 \frac{\delta}{\delta P^a_i} + \delta_\alpha^3 \frac{\delta}{\delta \xi} + \delta_\alpha^4 \frac{\delta}{\delta \pi_\xi} + \delta_\alpha^5 \frac{\delta}{\delta \chi} + \delta_\alpha^6 \frac{\delta}{\delta \pi_\chi} \right) \mathcal{C}_a \delta_\beta^9 -\left( -1 \right)^{\varepsilon_R\varepsilon_S}  \alpha \leftrightarrow \beta \right] \;.
\end{eqnarray}
The first-order pre-symplectic structure contains the following zero-modes
\begin{eqnarray}
	\left( \nu_{[6]i} \right)^\alpha &=& 
	\left( \begin{matrix}
	\frac{\delta \mathcal{G}_i}{\delta P^b_j} & -\frac{\delta \mathcal{G}_i}{\delta \mathcal{A}^j_b} & \frac{\delta \mathcal{G}_i}{\delta \pi_\xi} & \frac{\delta \mathcal{G}_i}{\delta \xi} & \frac{\delta \mathcal{G}_i}{\delta \pi_\chi} & \frac{\delta \mathcal{G}_i}{\delta \chi} & -\delta_i^j & 0 & 0^a
	\end{matrix} \right) \;, \label{eq::rotZM} \\
	\left( \nu_{[6]} \right)^\alpha &=& 
	\left( \begin{matrix}
	\frac{\delta \mathcal{C}}{\delta P^b_j} & -\frac{\delta \mathcal{C}}{\delta \mathcal{A}^j_b} & \frac{\delta \mathcal{C}}{\delta \pi_\xi} & \frac{\delta \mathcal{C}}{\delta \xi} & \frac{\delta \mathcal{C}}{\delta \pi_\chi} & \frac{\delta \mathcal{C}}{\delta \chi} & 0^j & -1 & 0^a
	\end{matrix} \right) \;, \label{eq::HamiltonZM} \\
	\left( \nu_{[7]a} \right)^\alpha &=& 
	\left( \begin{matrix}
	\frac{\delta \mathcal{C}_a}{\delta P^b_j} & -\frac{\delta \mathcal{C}_a}{\delta \mathcal{A}^j_b} & \frac{\delta \mathcal{C}_a}{\delta \pi_\xi} & \frac{\delta \mathcal{C}_a}{\delta \xi} & \frac{\delta \mathcal{C}_a}{\delta \pi_\chi} & \frac{\delta \mathcal{C}_a}{\delta \chi} & 0^j & 0 & -\delta^a_b
	\end{matrix} \right) \;. \label{eq::diffZM}
\end{eqnarray}

Since the first-order symplectic potential is equal to zero, we have that, according to Eq. (\ref{eq::consistCond}), none of the zero-modes generate new constraints and are, thus, generators of on-shell gauge transformations given by Eq. (\ref{eq::OnShellGaugeTransf}). It follows, for instance, for an infinitesimal parameter $\rho^i$, the gauge invariance expressions for the corrected Ashtekar-Barbero connection and its conjugate momenta, respectively, are
\begin{eqnarray}
	\delta_{\vec{\rho}} \mathcal{A}^i_a = \frac{\delta}{\delta P^a_i} \int_\Sigma d^3x \rho^i G_i = - \mathcal(D)_a^{\mathcal{A}} \rho^i \;,
\end{eqnarray}
\begin{eqnarray}
	\delta_{\vec{\rho}} P^a_i = -\frac{\delta}{\delta \mathcal{A}^i_a} \int_\Sigma d^3x \rho^i G_i = \epsilon_{ij}^{\ \ k} \rho^j P^a_k \;,
\end{eqnarray}
and
\begin{eqnarray} \label{eq::xiGaugeRotation}
	\delta_{\vec{\rho}} \xi = \frac{\delta}{\delta \pi_\xi} \int_\Sigma d^3x \rho^i G_i = \rho^i  \tau_i \xi  \;.
\end{eqnarray}
It is known\cite{Immirzi:1996dr} that Eq. (\ref{eq::xiGaugeRotation}) has the effect of local rotations in the field $\xi$.

For another distinct infinitesimal parameter $\varepsilon^a$, we have that the expression for the spatial diffeomorphism gauge transformation for the corrected Ashtekar-Barbero connection and its conjugate momenta are, respectively, 
\begin{eqnarray}
	\delta_{\vec{\varepsilon}} \mathcal{A}^i_a = \frac{\delta}{\delta P^a_i} \int_\Sigma d^3x \ \varepsilon^a \mathcal{C}_a = \varepsilon^b\mathcal{F}^{(\mathcal{A})i}_{\ \ ba} + \mathcal{D}^{(\mathcal{A})}_a \left( \varepsilon^c \mathcal{A}^i_a \right) = \mathcal{L}_{\vec{\varepsilon}} \mathcal{A}^i_a \;,
\end{eqnarray}
\begin{eqnarray}
	\delta_{\vec{\varepsilon}} P^a_i &=& -\frac{\delta}{\delta \mathcal{A}^i_a} \int_\Sigma d^3x \ \varepsilon^a \mathcal{C}_a = \int_\Sigma d^3x  \left[ 2\frac{\delta}{\delta \mathcal{A}^i_a} \left( \varepsilon^b P^c_j \partial_{[a} \mathcal{A}^j_{b]} \right) - \varepsilon^a \partial_c P^c_i \right] \nonumber \\
	&=&  \int_\Sigma d^3x  \left[  \varepsilon^c \partial_{c} P^a_i - P^b_i \partial_{b} \varepsilon^a + P^a_i \partial_b \varepsilon^b \right] \nonumber \\
	&=& \mathcal{L}_{\vec{\varepsilon}} P^a_i \label{eq::GaugeInvarPia} \;,
\end{eqnarray}
also
\begin{eqnarray}
	\delta_{\vec{\varepsilon}} \xi &=& \frac{\delta}{\delta \pi_\xi} \int_\Sigma d^3x \ \varepsilon^a \mathcal{C}_a =  \int_\Sigma d^3x  \frac{1}{2} \frac{\delta}{\delta \pi_\xi} \left( \varepsilon^a \pi_\xi \partial_a \xi - (\partial_a \pi_\xi) \varepsilon^a \xi \right) \\
	\delta_{\vec{\varepsilon}} \xi &=& \varepsilon^a \partial_a \xi + \frac{1}{2} \left( \partial_a \varepsilon^a \right) \xi = \mathcal{L}_{\vec{\varepsilon}} \xi  \;,
\end{eqnarray}
As has been obtained in \cite{Date:2011rd}.
	
The gauge transformations associated with the Hamiltonian constraint are much more involved mathematically, besides, its form depends on another set of transformations needed to deal with the problem in the context of LQG. Such development is beyond the scope of this work, but the result is presented in detail in \cite{Bojowald:2007nu}.
	
In order to count the number of the degrees of freedom we follow Eq. (\ref{eq::NDF}). In the last iteration we have $33$ field components from (Eq. (\ref{eq::sympVector01})), considering that $\psi$ and $\eta$ are bi-spinors, hence $\xi$, $\chi$ and their momenta account for two components each; from (Eqs. (\ref{eq::GaussConstraintNew}-\ref{eq::DiffConstraintNew})) we have $7$ uneliminated constraints; and, in the last iteration, $7$ gauge generating zero-modes were found (Eqs. (\ref{eq::rotZM}-\ref{eq::diffZM})), then the number of degrees of freedom of the theory is
\begin{equation}
	\mbox{NDF} = \frac{1}{2}(N^{(1)} - 2M - G) = \frac{1}{2}( 33 - 2 \times 7 - 7 ) = 6\;. 
\end{equation}
Essentially, there are two degrees of freedom relative to the gravitational theory classically equivalent to General Relativity\cite{Rodrigues:2018ioe} and four more degrees of freedom associated with the fermions\cite{Schwartz:2013pla}.
	
\section{Conclusion}

The work reviewed the canonical formulation of a non-perturbative quantum gravitational theory coupled to fermions in the context of a symplectic formalism. After asserting a linearized Lagrangian, it was straightforward, using the BW algorithm, to find the constraints of the theory and its pre-symplectic structure. In this process it was possible to highlight the problems associated with a particular ``natural'' choice of fermion coordinates and review how this problem can be dealt with using half-densities Grassmann variables introduced by Thiemmann\cite{Thiemann:1997rq,Thiemann:2007zz}, showing a different point of view of how the issue arises in the context of symplectic geometry. The second class constraints were then solved, reducing the symplectic space, leading to the pre-symplectic space composed only of gauge generating constraints. The gauge invariance expressions were obtained for the main fields of the theory. Finally, the counting of degrees of freedom was performed. In the present approach, finding the constraints and the reduction of the phase-space was performed in a systematic manner using the BW algortihm of the Faddeev-Jackiw formalism. 

\section{Acknowledgments}
I thank Davi C. Rodrigues for discussions on the symplectic method.

\appendix
\section{Symplectic analysis with Grassmann variables} \label{sec::MetSimpGrass}
In the following exposition we build upon the work of Govaerts\cite{Govaerts:1990mn}.
Consider a field system given by the action
\begin{eqnarray}
	S[\xi^\alpha] = \int_{t_1}^{t_2} dt L(\xi, \dot{\xi}) \;,
\end{eqnarray}
where the $\xi$'s are the fields of the system, and with Lagrangian given by 
\begin{eqnarray}
	L(\xi_a, \dot{\xi}_a) = \int_{\Sigma_t} d^3x \ \mathcal{L}( \xi^\alpha(x,t), \dot{\xi}^{\alpha}(x,t) ) \;.
\end{eqnarray}
The Grassmann partity of the coordinates $\xi^\alpha$ assume the values $\varepsilon^\alpha = 0$ for an odd variable and $\varepsilon^\alpha = 1$ for an even variable.

In order to apply the symplectic analysis in a system we suppose the existence of a Lagrangian linearized in the velocities that can be written as
\begin{eqnarray}\label{eq::0OLagrangian}
	L(\xi_a, \dot{\xi}_a) = \dot{\xi}^\alpha a_\alpha - V(\xi^\alpha) \;.
\end{eqnarray}
From which we obtain the pre=symplectic matrix
\begin{eqnarray}\label{eq::sympMatrix}
	f_{\alpha \beta} = \frac{\delta a_\beta}{\delta \xi^\alpha} - (-1)^{\varepsilon^\alpha \varepsilon^\beta} \frac{\delta a_\alpha}{\delta \xi^\alpha} \;.
\end{eqnarray}
By definition, the Grassmann parity of $f_{\alpha \beta}$ is equal to $(\varepsilon^\alpha + \varepsilon^\beta)$, from which we obtain
\begin{eqnarray}
	f_{\beta \alpha} = -(-1)^{\varepsilon_\alpha \varepsilon_\beta} f_{\alpha \beta} \;.
\end{eqnarray}
If $f_{\alpha \beta}$ is regular, the Euler-Lagrange equations yield the equations of motion
\begin{eqnarray}
	\dot{\xi^\beta} f_{\beta \alpha} = - \frac{\delta V}{\delta \xi^\alpha} \;,
\end{eqnarray}
or
\begin{eqnarray}
	(-1)^{\varepsilon_\beta} f_{\alpha \beta} \dot{\xi^\beta} = \frac{\delta V}{\delta \xi^\alpha} \;.
\end{eqnarray}
Inverting $f_{\alpha \beta}$, we have
\begin{eqnarray}
	\dot{\xi^\alpha} = -\frac{\delta V}{\delta \xi^\beta} \left( f^{-1} \right)^{\beta \alpha} = (-1)^{\varepsilon^\alpha}  \left( f^{-1} \right)^{\alpha \beta} \frac{\delta V}{\delta \xi^\beta} \;.
\end{eqnarray}
Noting that
\begin{eqnarray}
	\left( f^{-1} \right)^{\beta \alpha} = (-1)^{\varepsilon^\alpha + \varepsilon^\beta + \varepsilon^\alpha \varepsilon^\beta}  \left( f^{-1} \right)^{\alpha \beta} \;.
\end{eqnarray}
	
If $f_{\alpha \beta}$ is singular, there are $N$ (left) zero-modes $\nu^\alpha_{[A]}$, with $A=1,\ldots,N$, satisfying
\begin{eqnarray} 
	\int d^3x^\prime \ \nu^\alpha_{[A]}(x^\prime) f_{\alpha \beta}(x,x^\prime) = 0\;.
\end{eqnarray}
	
For right zero-modes, $\tilde{\nu}^\alpha_{[A]}$, we have
\begin{eqnarray}
	f_{\alpha \beta} \tilde{\nu}^\alpha_{[A]} &=& 0 \;, \\
	\tilde{\nu}^\alpha_{[A]} &=& -(-1)^{\varepsilon^\alpha} \nu^\alpha_{[A]} \;.
\end{eqnarray}
	
From the Euler-Lagrange equations, we have that the zero-modes must satisfy the constraint equations given by
\begin{eqnarray}\label{eq::consistCond}
	\omega_{[A] \beta}(x) = \nu^\alpha_{[A]}(x) \frac{\delta V}{\delta \xi^\alpha} = 0  
\end{eqnarray}
which must be either solved or imposed as constraints to the next-order Lagrangian in order to obtain a Lagrangian restricted to the constraints surface, or
\begin{eqnarray}
	L^\prime = L \Big|_{\omega = 0} = \dot{\xi}^\alpha a_\alpha + \dot{\lambda}^\alpha_{[A]} \omega_{[A] \alpha} - V \Big|_{\omega = 0}  \;.
\end{eqnarray}
In case there are $M$ gauge generating constraints, gauge conditions of the type 
\begin{eqnarray}
	C_{[B] \alpha} (\xi^\alpha) = 0 
\end{eqnarray}
must be imposed, with $B= 1,\ldots, M$, as additional kinetic terms in the Lagrangian,
\begin{eqnarray}
	L^{\prime \prime} = \dot{\xi}^\alpha a_\alpha + \dot{\lambda}^\alpha_{[A]} \omega_{[A] \alpha} + \dot{u}^\alpha_{[B]} C_{[B] \alpha}  - V \Big|_{\omega = 0}  \;.
\end{eqnarray}
Enabling, in this way, to write the inverse matrix of the found symplectic 2-form $f^{\ \prime \prime}_{\alpha \beta}$ and the equations of motion through the symplectic structure defined by means of generalized brackets
\begin{eqnarray}
	\{ \xi^\alpha, \xi^\beta \}^* = \left( f^{-1} \right)^{\alpha \beta} \;.
\end{eqnarray} 
	
Furthermore, the set of $M$ zero-modes generators of gauge transformations yield the on-shell gauge relations as
\begin{eqnarray}\label{eq::OnShellGaugeTransf}
	\delta_{\eta} \xi^\alpha = \nu^\alpha \eta \;.
\end{eqnarray}
	
To conclude, we enunciate the formula used to compute the number of degrees of freedom of a theory within the BW algorithm. The formula was derived for regular variables, but must hold also for Grassmann variables. Once the Lagrangians of all iterations must be equivalent. If the number of coordinates in the last, say $k$-th, iteration is $N^{(k)}$, and if $M$ unsolved constraints were found, then the number of degrees of freedom must be\cite{Rodrigues:2018ioe}
\begin{eqnarray}\label{eq::NDF}
	\mbox{NDF} = \frac{1}{2} \left( N^{(k)} - 2M - G \right) \;,
\end{eqnarray}
where $G$ is the number of independent zero-modes that do not lead to any new constraint.
	
\bibliographystyle{apsrev4-1}
\bibliography{SymplecticHolst-Dirac}
	
\end{document}